\documentclass{kais}

\usepackage{wrapfig,epsf}
\usepackage{epsfig}
\usepackage{graphicx}
\usepackage{amssymb}
\usepackage{algorithm}
\usepackage{algorithmicx}
\usepackage{algpseudocode}
\usepackage{subfigure}
\usepackage{epsfig}
\usepackage[dcucite]{harvard}
\usepackage{multirow}
\usepackage{amsmath}
\usepackage{rotating}
\usepackage{url}
\usepackage{booktabs}
\usepackage{color}
\usepackage{ulem}

\newtheorem{myDef}{Definition}

\pubyear{2000}
\pagerange{\pageref{firstpage}--\pageref{lastpage}}
\volume{xxx}

\begin{document}

\title{Mutual Clustering on Comparative Texts via Heterogeneous Information Networks}

\author[SZ. Wang et al]{Jianping Cao$^1$,Senzhang Wang$^{2*}$,Danyan Wen$^3$,Zhaohui Peng$^4$,Philip S. Yu$^{5,6}$,Fei-yue Wang$^{1,7,8,9}$\\
$^1$Research Center for Military Computational Experiments and Parallel Systems Technology, \\National University of Defense Technology, Changsha 410073, China;\\
$^{2*}$Corresponding author: Key Laboratory of Safety-Critical Software (Nanjing University of Aeronautics and Astronautics), \\Ministry of Industry and Information Technology, Nanjing 211106, China, szwang@nuaa.edu.cn; \\
$^3$School of Economics \& Management, Nanjing University of Science and Technology, \\Nanjing 210094, China, wendy2018@njust.edu.cn;\\
$^4$School of Computer Science and Technology, Shandong University, Qingdao, China, pzh@sdu.edu.cn;\\
$^5$Department of Computer Science, University of Illinois at Chicago, Chicago 60607, USA, psyu@uic.edu;\\
$^6$Institute for Data Science, Tsinghua University, Beijing 100084, China;\\
$^7$The State Key Laboratory of Management and Control for Complex Systems(SKL-MCCS),\\ Institute of Automation, Chinese Academy of Sciences(CASIA), Beijing, China;\\
$^8$Qingdao Academy of Intelligent Industries, Qingdao 266000, China;\\
$^9$Institute of Systems Engineering, Macau University of Science and Technology, Macau 999078, China.
}
\maketitle

\begin{abstract}
Currently, many intelligence systems contain the texts from multi-sources, e.g., bulletin board system (BBS) posts, tweets and news. These texts can be ``comparative'' since they may be semantically correlated and thus provide us with different perspectives toward the same topics or events. To better organize the multi-sourced texts and obtain more comprehensive knowledge, we propose to study the novel problem of Mutual Clustering on Comparative Texts (MCCT), which aims to cluster the comparative texts simultaneously and collaboratively. The MCCT problem is difficult to address because 1) comparative texts usually present different data formats and structures and thus they are hard to organize, and 2) there lacks an effective method to connect the semantically correlated comparative texts to facilitate clustering them in an unified way. To this aim, in this paper we propose a Heterogeneous Information Network-based Text clustering framework HINT. HINT first models multi-sourced texts (e.g. news and tweets) as heterogeneous information networks by introducing the shared ``anchor texts'' to connect the comparative texts. Next, two similarity matrices based on HINT as well as a transition matrix for cross-text-source knowledge transfer are constructed. Comparative texts clustering are then conducted by utilizing the constructed matrices. Finally, a mutual clustering algorithm is also proposed to further unify the separate clustering results of the comparative texts by introducing a clustering consistency constraint. We conduct extensive experimental on three tweets-news datasets, and the results demonstrate the effectiveness and robustness of the proposed method in addressing the MCCT problem.
\end{abstract}

\section{Introduction}
\label{sec:sec1}

\begin{figure*}
\centering
\epsfig{file=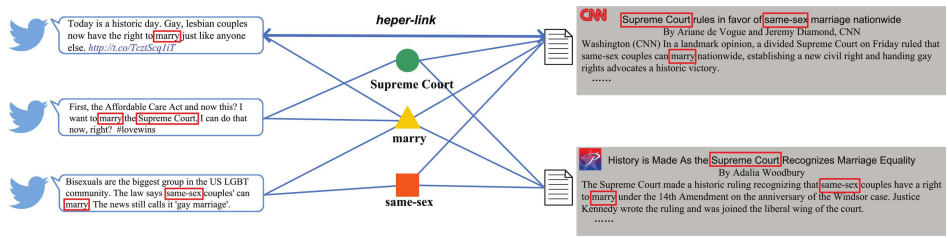, height=1.2in}
\caption{An illustration of the comparative texts. Tweets and news are discussing the same topic and they can be referred as ``comparative texts''. We use the hyper-links, common words and entities to connect the comparative texts.}
\label{fig:fig11}
\end{figure*}

With the booming of Web 2.0, text information of different sources increases at an unprecedented rate. Such texts can come from various types of websites and present different formats or structures, but in many occasions they are related to the same events or talking about the same topics as shown in \ref{fig:fig11}. Following the definition given in \cite{zhai2004cross}, these semantically correlated texts of different sources are referred as Comparative Texts (CTs), e.g. tweets from Twitter and news articles from news websites that are both related to U. S. president Donald Trump. Comparative texts can usually provide us with more comprehensive information compared to the single-sourced texts. For example, news articles tell us the fact and official views of an event, while tweets reveal personal affections and attitudes toward the event \cite{gao2012joint,kais2016}. Therefore, analyzing the comparative texts collaboratively rather than isolatedly is strongly necessary to help us better organize text information and obtain more comprehensive knowledge.

To this aim, we study the novel problem of Mutual Clustering on Comparative Texts (MCCT), and more specifically we focus on clustering the comparative texts with heterogeneous structures simultaneously and collaboratively. The MCCT problem is different from the Comparative Text Mining (CTM) proposed \cite{zhai2004cross} which assumes the comparative texts are of the same type or at least with homogeneous structures. In MCCT, the texts are of different types and could be heterogeneous in text formats, style, or structures. For example, news are typically well-crafted and fact-oriented long stories written in more normal texts, while tweets are mostly personalized and free-style short texts \cite{gao2012joint}. The MCCT problem is also quite different from the works of clustering short texts \cite{guo2013linking,sahami2006web}, which uses news or other text data as auxiliary information to facilitate short text clustering like tweets clustering. A major difference is that these works only focus on clustering one type of texts, and consider the other types of texts as auxiliary information. Different from previous works, the proposed MCCT problem consists of the following two tasks: (1) clustering the comparative text collections simultaneously in order to better organize them into correlated topics, and (2) connecting the semantically correlated clusters together to get a comparative cluster results for each topic. For example, given a collection of tweets and news that both contain the texts related to the topic of \textit{\#Obamacare}, MCCT aims to discover the semantically correlated clusters of \textit{\#Obamacare} from the two types of texts collaboratively, and connect them at the same time.

The MCCT problem is difficult to address due to the following two major challenges. First, the formats and structures of comparative texts can be quite different, leading to the difficulty of using traditional methods to represent and organize them. Second, it is also difficult to mine the semantic correlations among comparative texts to effectively connect the texts of different types and clustering them simultaneously and collaboratively. To address the above challenges, in this paper we propose a Heterogeneous Information Network-based Text clustering (HINT) framework for mutually clustering on comparative text. As Heterogeneous Information Networks (HINs) can model multiple types of objects and reflect the complex relations among them \cite{sun2009RankClus,wang2015incorporating,Zhang2016Text,PAKDD2016}, they are very suitable to represent the comparative texts and their semantic connections. Meanwhile, we can also use both the lexical and structural information of HINs to guide the mutual clustering on the comparative texts. Specifically, the proposed HINT framework consists of the following three steps. First, we extract the objects like keywords and named entities from the texts, based on which we construct two heterogeneous information networks for the two types of tweets, respectively. To bridge the semantic gap between the two types of texts, we align parts of the texts in the tweets with news based on their similar semantic meanings and refer to these texts as ``anchored texts.'' Next, we construct the similarity matrices for both types of texts based on the constructed HINs, and also construct a transition matrix to transfer knowledge among the texts of different types. Finally, we cast the problem into a spectral clustering on the three matrices with a non-convex objective function. A curvilinear search algorithm is further proposed to effectively solve the non-convex optimization problem.
The primary contributions of this paper can be summarized as follows:

$\bullet$ We study the novel problem of mutual clustering on the comparative texts (MCCT) w.r.t. their semantic correlations. To the best of our knowledge, this is the first attempt to mutually cluster multi-sourced texts with different formats and structures.

$\bullet$ We utilize HINs to model and organize comparative texts, and propose a HINT model to address the MCCT problem. HINT casts MCCT into an optimization problem with a non-convex objective function.

$\bullet$ We propose a mutual clustering algorithm under the HINT framework, which employs a curvilinear search method to iteratively seek the solution of the non-convex optimization problem.

$\bullet$ Extensive experiments are conducted on three tweets-news comparative text datasets, and the experimental results show the effectiveness of the proposed HINT.

The remainder of this paper is organized as follows. The related work of the study is presented in section 2. Section 3 will give a formal definition of the studied problem. In section 4, we will present the details of the proposed HINT model. Next, a matrix-based mutual clustering method on comparative texts will be presented in section 5. We show the experimental results in section 6, and finally conclude this paper in section 7.

\section{Related Work}\label{sec:sec6}

As an important research topic in text analysis \cite{Tian2017Dynamic}, comparative text clustering studied in this paper aims to mutually analyze different types of texts by taking their semantic correlations into account. This work is closely related to the topics of text representation, graph-based text processing, clustering on HINs and the co-clustering of texts. In this section, we will introduce the related works from the four aspects and discern our problem from them.

\textbf{Text Representation.} As the preliminary step of most text mining tasks, how to represent texts is one of the key challenges in the field of text mining \cite{Han2011Data}. The typical methods in this area are based on bag-of-words (BoW) \cite{Mladenic1998word}, such as TF-IDF \cite{Salton1988Term}, $n$-grams model \cite{Keselj2003N}, and vector-space model \cite{erk2008structured}. Mladenic and Grobelnik \cite{erk2008structured} for the first time proposed an efficient algorithm for generation of new features to enrich the BoW document representation. Keselj et al. \cite{Keselj2003N} proposed to use byte-level $n$-grams to generate features from the texts of author profiles. Salton and Buckley \cite{Salton1988Term} studied the problem that the assignment of appropriately weighted single terms can obtain better text representation features and produce better retrieval results. Erk and Pado \cite{erk2008structured} presented a structure vector space model to incorporate the selectional preferences for words' argument positions. As for the representation of tweet-like short texts \cite{neuro2015}, except for the above mentioned methods, some previous works also applied HINs to represent various networks with text information including the networks, the re-tweet networks \cite{yan2012tweet,kdd14}, or term-tweet correlation networks \cite{hua2013sted}. For example, Yan et al. \cite{yan2012tweet} proposed to model the tweets of their authors as a heterogenous network to co-rank the the two types of entires simultaneously for recommendation. However, previous text representation methods mainly focus on only one type of texts, which is quite different from our study that makes efforts to bridge the gap across two types of texts.

\textbf{Graph-based Text Processing.} The graph-based text processing models enhances traditional text processing models by incorporating some structural features into the documents, e.g., the ordering of words \cite{Aggarwal2013Towards} and the connection between authors and publications \cite{Villarreal2016Local}. The core idea of graph-based text processing is to construct a graph by considering words or documents as nodes, and then run some data processing algorithms on the graph via topological methods \cite{mooney2007learning,wei2010A,Aggarwal2013Towards}. Mooney \cite{mooney2007learning} studied the task of semantic parsing which aimed to map a natural language sentence into a complete, formal meaning representation. The semantic parsers were developed with machine learning algorithms that were trained on the constructed graph of sentences pairs. Wei et al. \cite{wei2010A} presented a novel document-sensitive graph model for multi-document summarization. The proposed model emphasized the influence of global document set information on local sentence evaluation. Aggarwal and Zhao \cite{Aggarwal2013Towards} introduced the concept of \textit{distance graph representations} of text data. Such representations preserved information about the relative ordering and distance between the words in the graphs and provided a much richer representation in terms of sentence structure of the underlying data. Recent related work mainly focused on the construction of suitable graph-based models for some specific tasks \cite{pinto2014graph,Balachandran2012Interpretable}. In this field, HINs are often used to enhance the semantic meanings \cite{yan2012tweet}. Almost all of these works focused on the problem of clustering one type of documents, and thus they are not suitable to be applied to our comparative texts clustering task.

\textbf{Clustering on HINs.} The problem of clustering on heterogeneous information networks (HINs) has attracted rising research interests recently \cite{PAKDD2016}. Sun et al. for the first time proposed to use ``meta-path'' for the calculation of item similarities in HINs \cite{sun2009RankClus}. Due to the complexity of HINs, different researchers may focus on different aspects of the HINs, and thus the similarity measure between the same pair of items can be various \cite{WAIM}. Therefore, researchers also proposed to cluster HINs under the guidance provided by the users \cite{Sun2013PathSelClus,sun2012Relation,PAKDD2015}. Sun et al. \cite{kdd2012} studied how to leverage the rich semantic meaning of structural types of objects and links in HINs, and developed a structural analysis approach on mining semi-structured, multi-typed heterogeneous information networks. They designed a probabilistic model which cluster the objects of different types in HINs into a common hidden space, by using a user-specified set of attributes, as well as the links from different relations \cite{sun2012Relation}. Sun et al. \cite{Sun2013PathSelClus} further studied how to use \textit{meta-path} to control clustering with distinct semantics in HINs. Since the HINs  carry the type information about entities and relations, they have been widely used to represent texts for text mining tasks \cite{Zhang2016Text}. Li et al. \cite{www2017} studied the problem of clustering objects in an Attributed HIN. They took into account objects' similarities with respect to both object attribute values and their structural connectedness in the network. Zhang et al. \cite{ICDM2016}  proposed a new HIN clustering algorithm which can handle general HINs, simultaneously generated clusters for all types of objects, and used the similarity information of the same type of objects. Shi et al. have provided a comprehensive survey on heterogeneous information network mining \cite{TKDE2017}. One can refer to it for more details on exploring HINs for various data mining tasks.

\textbf{Co-clustering.} Co-clustering, or bi-clustering is the problem of simultaneously clustering rows and columns of a data matrix, which is often used to study the problems of clustering two different types of objects \cite{dhillon2003information,Shaham2012Sleeved}, e.g., genes and experimental conditions in bioinformatics \cite{Cheng2000Biclustering,Cho2004Minimum}, documents and words in text mining \cite{dhillon2003information}, and users and movies in recommender systems, etc. The co-clustering model can also be used in the study of transfer-learning. For example, it has been used to enrich the short text like tweets \cite{guo2013linking,jin2011transferring}.  Among these related works, one typical application of Co-clustering is on the clustering of documents and words in text mining \cite{dhillon2001co,Yan2013semi}. Guo et al. \cite{guo2013linking} proposed a tweets-News co-clustering method. However, different from our work, this work mainly focused on utilizing News documents to enrich the short tweets to better understand the semantics of short texts. Yan et al. \cite{Yan2013semi} proposed a new semi-supervised fuzzy co-clustering algorithm called SS-FCC for categorization of large web documents. SS-FCC was formulated as the problem of maximizing a competitive agglomeration cost function with fuzzy terms, taking into account the provided domain knowledge. In terms of graph, the co-clustering method focuses on the row and column of the bipartite graph matrix \cite{Cheng2016HICC}. Nie et al. \cite{NIPS2017} proposed a novel co-clustering method to learn a bipartite graph with $k$ connected components, where k is the number of clusters. The new bipartite graph learned in the model approximated the original graph but maintained an explicit cluster structure. Li et al. \cite{kais2015} proposed a generative model named author-topic-community (ATC) model for representing a corpus of linked documents. For each author, ATC model can dicover his/her topics and the communities the author belonging to through a generative co-clustering model. Since the clustering objects in our paper are semantically correlated rather than directly connected, it is hard to simply correlate the two types of objects, the assumption of co-clustering is not suitable for the studied problem.

\section{Preliminary and Problem Definition}\label{sec:sec2}

In this section, we first give two definitions to help us state the studied problem, and then we will formally define the proposed MCCT problem.
\begin{myDef}
HINs for Tweet and News. A tweet or news article can be represented in such a heterogeneous information network $\mathcal{G}(\mathcal{V},\mathcal{E})$, where vertexes $\mathcal{V}$ contain objects with the following types, text $(T)$,  word ($\mathcal{O}_w$), named entities ($\mathcal{O}_e$), and other types like mention ($\mathcal{O}_m$) and hashtag ($\mathcal{O}_h$) (only for tweet); and edges $\mathcal{E}$ represent the multiple types of relations connecting the tweet objects. Note that to discern the two types of texts, we use superscript ``(1)'' and ``(2)'', e.g. $T^{(1)}$ and $T^{(2)}$, to represent tweets, news and corresponding objects respectively.
\end{myDef}

Since many tweets may have hyper-links directing to other web pages like news sites, we refer these text pairs as $anchor~texts$.

\begin{myDef}
Anchor texts. Given a tweet $t_i^{(1)} \in T^{(1)}$ and a news article $t_i^{(2)} \in T^{(2)}$, if there is a hyper-link in $t_i^{(1)}$ directing to $t_i^{(2)}$ and they share some common objects such as words or entities, the two semantically correlated texts are referred to a pair of ``anchor texts,'' denoted as $(t_i^{(1)},t_j^{(2)})$ or $(t_j^{(2)},t_i^{(1)})$. All these ``anchor texts'' pairs form an anchor text collection $\mathcal{R}$.
\end{myDef}

The MCCT problem targets at detecting semantically correlated clusters $C^{(1)}$ and $C^{(2)}$ from tweet $T^{(1)}$ and news articles $T^{(2)}$ simultaneously, and it can be formally defined as:

\textit{Given two types of comparative texts $T^{(1)}$ and $T^{(2)}$ and the corresponding anchor text collection $\mathcal{R}$, the mutual clustering problem aims to obtain clusters $\{C^{(1)},C^{(2)}\}$ for $\{T^{(1)},T^{(2)}\}$ simultaneously, where each cluster $C_i^{(1)}$ in $C^{(1)}$  is semantically correlated to $C_j^{(2)}$ in $C^{(2)}$ , $i=1,\cdots,k^{(1)}$, $j=1,\cdots,k^{(2)}$.}

\section{Modeling Comparative Texts with HINs}\label{sec:sec3}

The proposed framework HINT consists of two major steps: the construction of similarity matrices and transition matrix based on the HINs for the mutual texts, and the mutual clustering approach via the similarity matrices and transition matrix. This section will introduce the first step of the framework. We will first introduce how to model tweets and news as HINs. Then we will calculate the similarities between the two kinds of texts and align the two types of texts.

\subsection{The Workflow of HINT}

\begin{figure*}
\centering
\epsfig{file=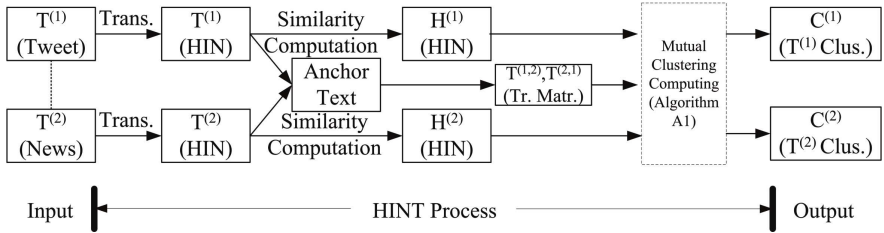, height=1.2in}
\caption{The workflow of our proposed framework.}
\label{fig:fig2-3}
\end{figure*}

The workflow of HINT is illustrated in Fig. \ref{fig:fig2-3}. One can see that the mutual clustering on the comparative texts are partially connected by some anchor text links. Our aim is to obtain two sets of text clusters for tweets and news respectively and connect two clusters that share similar semantic meanings. To this end, we propose HINT to cluster the comparative texts simultaneously and collaboratively.

The main steps of the HINT framework can be summarized as follows. First, we model the two types of texts as heterogeneous information networks (HINs). Meanwhile, the anchor texts that directly connected across the two types of comparative texts are selected for the guidance of mutual clustering. Then the initial confidence matrices $H^{(1)}$ and $H^{(2)}$ are calculated based on the HINs, and the transition matrices $T$ of anchor texts that contain the semantic correlation information are also constructed. The clustering problem is transformed to a constrained optimization problem, which can be solved by utilizing the confidence matrices and the transition matrix. The output of the HINT model is two clustering matrices for tweets and news  that contain the clustering information. We can use them to get the correlated clusters of the two types of text: tweets and news. The HINT model can further connect two clusters from the two cluster sets that share similar semantic meanings. For example, our method may connect the $i$-$th$ cluster of tweets with the $j$-$th$ cluster of news because both clusters are related to U.S. president Donald Trump. In the following subsection, we will give more detailed explanation of our framework.

\subsection{Transforming Comparative Texts to HINs}

\begin{figure*}
\centering
\epsfig{file=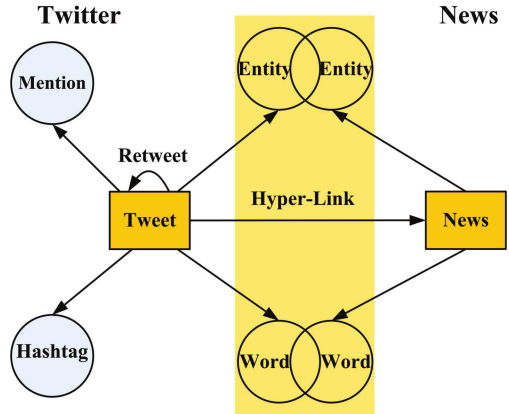, height=2.2in}
\caption{Schema of using HINs to model comparative texts.}
\label{fig:fig3}
\end{figure*}

To better represent the semantic meanings of the texts and effectively organize comparative texts, we transform comparative texts into HINs and connect them by anchor texts as shown in Fig. \ref{fig:fig3}.

\textbf{Texts to HINs.} A tweet can be modeled as a HIN with the following four types of objects as shown in the left part of Fig. \ref{fig:fig3}: words $\mathcal(O_w^{(1)})$, hashtags $\mathcal(O_h)$, mentions $\mathcal(O_m)$, and named entities $\mathcal(O_e^{(1)})$. Similarly, a news article can be modeled as a HIN with the following two types of objects as shown in the right part of Fig. \ref{fig:fig3}: words $\mathcal(O_w^{(2)})$, and named entities $\mathcal(O_e^{(2)})$. The topological structure of tweet and news information network is shown in Figure \ref{fig:fig3}, which forms two star network schemas, where the tweet and news are in the centers and all other objects are linked via them respectively. Links between the objects denote the semantical correlation among them, which will contribute to the similarity calculation of the texts. Note that the weight of an object is defined as $c$ if it appears $c$ times.

We extract the entities from tweets or news using the tools developed by \cite{manning2014stanford}. The named entities are extracted from ``words,'' ``hashtags (\#),'' and ``mentions ($@$)'' of the tweet. To more effectively extract the representative words of news, we utilize both the title and LDA \cite{Blei2003Latent} topic words to obtain the identical words in this study.

\textbf{Anchor Texts.} In order to get a consensus clustering, we correlate the two types of text via ``anchor texts.'' Previously, the phrase ``anchor text'' is used to describe ``the visible, click-able text in a hyperlink.'' In reality, a considerable amount of tweets contain hyper-links directing to other web pages like news sites. The source text (tweet) and target text (news) are usually correlated by such links with respect to semantic meanings. With this observation, and for the purpose of information transfer among the comparative texts,  we ``align'' these tweets with news articles in terms of semantic meanings and refer to these texts as anchored texts.

We first obtain the anchor texts through linked tweets and news pairs following \cite{guo2013linking}. Since there are still some tweets (about 7.9\% in our dataset) which are not closely related to the semantic meaning of news, we filter the anchor texts by common entities $(\mathcal{O}_e)$ and words $(\mathcal{O}_w)$. In this way, we develop a reliable correlation network across the two types of texts for further processing.

\subsection{Similarity and Transition Matrices Construction}
With the constructed network schema shown in Fig. \ref{fig:fig3}, we can build the similarity matrix through extracting various meta-paths  \cite{sun2011pathsim} to measure the semantic similarity among the texts. A meta-path is a sequence of relations among the objects of different types, which defines a new composite relation between the starting object type and the ending object type \cite{sun2011pathsim}. For example, the sequence $News \xrightarrow{contains} Entity \xleftarrow{contains} News$ is a meta-path which represents the relation of two news articles both containing an entity. We can see that both of the starting and ending types are $News$, and such a meta-path connects two $News$ sharing the same entity together for the sake of calculating their similarity. With the constructed network schema shown in Fig. \ref{fig:fig3}, we can build the similarity matrix incorporating various meta-paths information as constraints to measure the semantic similarity among the texts.

\textbf{Meta-paths of Tweet.}

$\bullet$ \textit{Retweets:} $Tweet \xrightarrow{retweet}  Tweet$ , whose notation is $T^{(1)} \leftrightarrow T^{(1)}$.

$\bullet$ \textit{Common Retweets:} $Tweet \xrightarrow{retweet} Tweet \xleftarrow{retweet} Tweet$, whose notation is  $T^{(1)} \rightarrow T^{(1)} \leftarrow T^{(1)}$.

$\bullet$ \textit{Common Objects:} $Tweet \xrightarrow{contains} Word/Entity/Mention/Hashtag/Hyper-\\Link \xleftarrow{contains} Tweet$, whose notation is  $T^{(1)} \rightarrow \mathcal{O}_w^{(1)}/\mathcal{O}_e^{(1)}/\mathcal{O}_m/\mathcal{O}_h/\mathcal{O}_l \leftarrow T^{(1)}$.

\textbf{Meta-paths of News.}

$\bullet$ \textit{Common Words:}  $News \xrightarrow{contains} Word \xleftarrow{contains} News$, whose notation is $T^{(2)} \rightarrow \mathcal{O}_w^{(2)} \leftarrow T^{(2)}$.

$\bullet$ \textit{Common Entities:}  $News \xrightarrow{contains} Entity \xleftarrow{contains} News$, whose notation is $T^{(2)} \rightarrow \mathcal{O}_e^{(2)}  \leftarrow T^{(2)}$.

Note that the named entities can contain multiple types (e.g., person, location, and organization names), and we calculate the entity similarity based on the named entity types given in \cite{wang2015incorporating}.

Note that the named entities can contain multiple types (e.g., person, location, and organization names), and we calculate the entity similarity based on the named entity types given in \cite{wang2015incorporating}.

$PathSim$ is an effective meta-path based similarity measurement \cite{sun2011pathsim}. Following the work \cite{sun2011pathsim}, we introduce a meta-path based similarity measure \textit{HINT similarity} defined as follows to calculate the comparative texts similarity.

\begin{myDef}
HINT Similarity. Let $P_i(x \leadsto y)$ and $P_i(x \leadsto \bullet)$ be the sets of path instances of $HINT_i$ going from node $x$ to $y$ and those going from $x$  to other nodes in the network. The semantic similarity between the two text nodes can be defined as follows,
\begin{equation}
Sim(x,y)=\sum_{i}w_i\left(\frac{|P_i(x \leadsto y)|+|P_i(y \leadsto x)|}{|P_i(x \leadsto \bullet)|+|P_i(y \leadsto \bullet)|} \right),
\end{equation}
\noindent where $w_i$ is the weight of the $i$-th meta-path, and we have $\sum_{i}w_i = 1$.
\end{myDef}

With the above definition, we construct the tweet-news similarity matrices whose dimensions are determined by the numbers of tweets and news. Let $A_i$  be the adjacency matrix of a type of texts with respect to the $i$-th meta-path. $A_i(m,n)=k$  denotes that there are $k$ concrete path instances between nodes $m$ and $n$ corresponding to the $i$-th meta-path. Then, the similarity matrix among the texts can be represented as:

\begin{equation}
S=\sum_{i}w_iS_i=\sum_iw_i \cdot Norm(A_i+ A_i^T),
\end{equation}

\noindent where $Norm(\cdot)$ is the normalization of matrix. Thus, the similarity matrices of all possible connections among tweets or news are constructed.

To use the anchor texts for information transfer between comparative texts, we formulate their semantic correlations as a transition matrix.

\begin{myDef}
Transition matrix. The transition matrix $T^{(1,2)}$ (or $T^{(2,1)}$) is a matrix representation of anchor texts with each element $T_{ij}^{(1,2)}=1$ denoting $(t_i^{(1)},t_j^{(2)}) \in \mathcal{R}$  and 0 otherwise. Here $t_i^{(1)} \in T^{(1)}$, $t_j^{(2)} \in T^{(2)}$ and $\mathcal{R}$ is the anchor text collection.
\end{myDef}

\begin{figure}
\centering
\epsfig{file=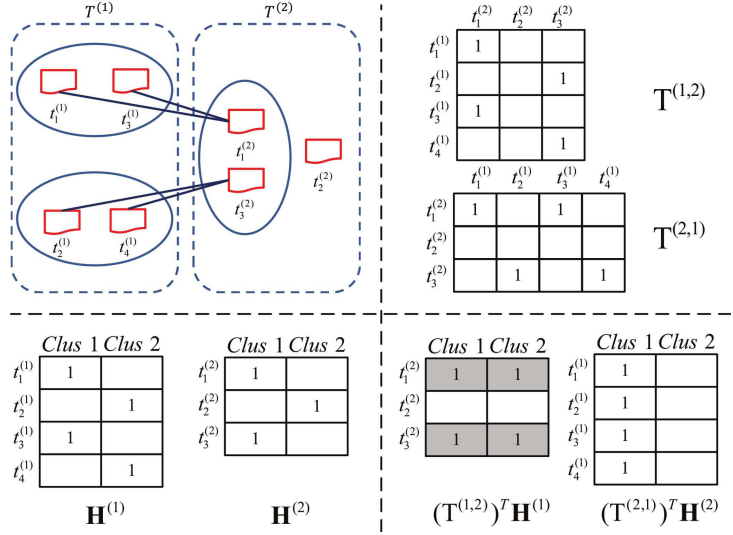, height=2.8in}
\caption{An illustration of the connections  between the two types of comparative texts (Top-left); transition matrices between the two networks (Top-right); clustering confidence matrices (Bottom-left); transmitted confidence matrices (Bottom-right).}
\label{fig:fig4}
\end{figure}

In this paper, we assume that a news article can be associated to many tweets, while a tweet is normally only associated to only one news article that is most related to it. Figure \ref{fig:fig4} gives an illustration of such ``one-to-many'' relationships. One can see that a piece of news $t_1^{(2)}$ (or $t_3^{(2)}$) is related to multiple tweets $t_1^{(1)}$,$t_3^{(1)}$ (or $t_2^{(1)}$,$t_4^{(1)}$), while each tweet is only related to one news article. Thus, in the transition matrix there can be multiple entries in each row with the value 1, while only one entry in each column with the value 1. Note that if the tweets are allowed to be related with more than one news, that is the relationships between tweets and news texts become $n$-to-$n$. The constraints between the two types of texts will be loosening, causing some unrelated texts to be connected after the transmitting operation, and leading to the confidence matrix out of control. Therefore,  the relationships between tweets and news articles used in this paper are subject to ``1-to-n.''

\section{Mutual Clustering on Comparative Texts}\label{sec:sec4}
In this section, we introduce the mutual clustering algorithm which contains three major steps. Firstly, we use spectral clustering algorithm to cluster tweets and news articles separately. Then, we introduce the inconsistency penalty (explain later) and use it as a constraint to refine the two clustering results. Finally, we propose a balanced clustering approach between tweets and news. As the objective function is non-convex, we also propose to solve it by applying a curvilinear searching approach.

\subsection{Clustering Comparative Texts}

We apply a partitioning method on tweet and news articles similarity matrices which divide the similarity network into several partitions by cutting connecting edges among them. The cutting activities need to balance the inner traits and outer traits. Therefore, cutting connections in a network will lead to some costs inevitably. Theoretically, the optimal clustering result can be achieved by minimizing the cost. For the convenience of comparison across networks, we adopt the normalized cut \cite{shi2000normalized,zhang2014mutual} as the cost criteria of partitioning or clustering in this paper. For all texts in $T^{(1)}$  and $T^{(2)}$, the partitioning result can be reported by the confidence matrix $\mathbf{H}$, where $\mathbf{H}=[h_1,h_2,\cdots,h_n]^T$ , $n=|T|$ , $h_i=(h_{i,1},h_{i,2},\cdots,h_{i,k})$  and $h_{ij}$  denotes the confidence that $t_i\in T$  is in cluster $C_j\in C$. The optimal $\bf{H}$  that minimizes the normalized cut cost can be obtained by solving the following objective function:

\begin{equation}
\begin{split}
\min_{\mathbf{H}} Tr(\mathbf{H}^T\mathbf{L}\mathbf{H}),\\
s.t. ~\mathbf{H}^T\mathbf{L}\mathbf{H} =\mathbf{I},
\label{eq:eq3}
\end{split}
\end{equation}

\noindent where $\mathbf{L}=\mathbf{D}-\mathbf{S}$. The diagonal matrix $\mathbf{D}$ is a matrix for the normalization of $\mathbf{S}$, where $\mathbf{D}(i,i) = \sum_j \mathbf{S}(i,j)$  on its diagonal, and $\mathbf{I}$ is an identity matrix.

\subsection{Penalty on Mutual Clustering Inconsistency}

The uniqueness of different types of texts could potentially lead to inconsistency between the final clustering results on the two types of texts. One typical case of the inconsistency is that two tweets correlated to the same news article are partitioned into different clusters. In such a case, a penalty should be assigned to avoid such an inconsistency. Another case is, two pieces of news articles are partitioned into the same cluster, but the tweets in their corresponding anchor texts are partitioned into different clusters. In this case, a penalty should also be assigned into the objective function as we aim to obtain a consensus clustering result across the two types of texts. By minimizing such inconsistency during clustering, we further refine the clustering results.

In this paper, we denote the clustering results of $T^{(1)}$ and $T^{(2)}$ as $C^{(1)} = \{C_1^{(1)}, C_2^{(1)}, \cdots, C_{k^{(1)}}^{(1)}\}$ and $C^{(2)} = \{C_1^{(2)}, C_2^{(2)}, \cdots, C_{k^{(2)}}^{(2)}\}$, respectively. Supposing there are two news articles $t_i^{(2)}$ and $t_j^{(2)}$ from $T^{(2)}$, and they are anchored to $t_{i_1,\cdots, i_m}^{(1)}$ and $t_{j_1,\cdots, j_n}^{(1)}$, respectively. Note that, here we suppose the corresponding relationships between news and tweets are 1-N. The confidence scores of them being partitioned into $k^{(2)}$ clusters are $\mathbf{h}_i^{(2)}$ and $\mathbf{h}_j^{(2)}$, respectively. Similarly, the corresponding correlated tweets $t_{i_1,\cdots,i_m}^{(1)}$ and $t_{j_1,\cdots,j_n}^{(1)}$ in $T^{(1)}$ being partitioned into $k^{(1)}$ clusters are $\mathbf{h}_{i_1,\cdots,i_m}^{(1)}$ and $\mathbf{h}_{j_1,\cdots,j_n}^{(1)}$, respectively. Then, the clustering  \textit{inconsistency} can be defined as follows.

\begin{myDef}
\textit{Mutual Clustering Inconsistency.} Mutual clustering inconsistency is defined as the difference of the confidence scores that $t_i^{(2)}$ and $t_j^{(2)}$ are partitioned into the same cluster in $T^{(2)}$ (denoted as $\mathbf{h}_i^{(2)}(\mathbf{h}_j^{(2)})^T$), and the scores that corresponding tweets $t_{i_1,\cdots,i_m}^{(1)}$ and $t_{j_1,\cdots,j_n}^{(1)}$ are partitioned into the same cluster in  $T^{(1)}$ (denoted as $\mathbf{h}_{i_p}^{(1)}(\mathbf{h}_{j_q}^{(1)})^T)^2 (p=1,\cdots,m;~q=1,\cdots,n)$).  Formally, their inconsistency can be calculated as follows:
\begin{equation}
d_{t_i^{(2)}t_j^{(2)}} = \sum_{p=1}^m\sum_{q=1}^n(\mathbf{h}_i^{(2)}(\mathbf{h}_j^{(2)})^T - \mathbf{h}_{i_p}^{(1)}(\mathbf{h}_{j_q}^{(1)})^T)^2.
\end{equation}
\end{myDef}

If one of the news texts $t_i^{(2)}$ and $t_j^{(2)}$ is non-anchored text, their inconsistency $d_{t_i^{(2)}t_j^{(2)}}$ should be $0$. Formally, the total inconsistency of the clustering results and the normalized total inconsistency can be defined as:

\begin{equation}
d(C^{(1)},C^{(2)})= \sum_{i=1}^{n^{(2)}}\sum_{j=1}^{n^{(2)}}d_{t_i^{(2)}t_j^{(2)}},
\label{eq:eq5}
\end{equation}

\begin{equation}
Nd(C^{(1)},C^{(2)})=\frac{d(C^{(1)},C^{(2)})}{|\mathcal{R}|(|\mathcal{R}|-1)},
\label{eq:eq6}
\end{equation}

\noindent where $n^{(2)}=|T^{(2)}|$, and $|\mathcal{R}|$ refers to the total number of anchor texts. The normalization of inconsistency makes it independent of the number of anchor texts. Moreover, such a normalization can prevent it from favoring highly consented clustering results when the anchor texts are abundant but have no significant effect on the dataset where the anchor texts are rare. The normalization is significantly different from the absolute clustering discrepancy cost used in \cite{cheng2013flexible}.

Therefore, in order to get the optimal consensus clustering results of $T^{(1)}$  and $T^{(2)}$, we expect to obtain
$\hat{C}^{(1)}$, $\hat{C}^{(2)}$ as follows,

\begin{equation}
\hat{C}^{(1)},\hat{C}^{(2)}=arg\min_{C^{(1)},C^{(2)}}Nd(C^{(1)},C^{(2)}).
\end{equation}

The normalized inconsistency objective function can also be represented by the clustering result confidence matrices  $\mathbf{H}^{(1)}$  and $\mathbf{H}^{(2)}$. Since the two types of texts are partially ``anchored,'' and the non-anchored texts are not involved in the calculation of inconsistency, we have to filter the non-anchored texts first in the calculation of  $\mathbf{H}^{(1)}$  and $\mathbf{H}^{(2)}$. Here we use the anchor transition matrix to prune the results of non-anchored texts. Let  $\bar{\mathbf{H}}^{(1)}=(T^{(1,2)})^T\mathbf{H}^{(1)}$ represents the clustering information of $C^{(1)}$  transmitting to $T^{(2)}$, and $\bar{\mathbf{H}}^{(2)}=(T^{(1,2)})^T(T^{(2,1)})^T\mathbf{H}^{(2)}$ represents the clustering information of $C^{(1)}$  transmitting to  $T^{(2)}$. The matrix $\bar{\mathbf{H}}(\bar{\mathbf{H}})^T$  indicates whether pairs of ``anchors'' are belonging to the same cluster or not. The bottom right figure in Figure \ref{fig:fig4} illustrates the transition of the confidence matrices, and one can see that the clustering information is transmitted across the two types of texts. Note that after transmitting information, the news belong to two clusters, which is just a reflection of the tweet clustering, and the differences should be punished. Here, due to the mapping relationship is ``one-to-many,'' we map the clustering information of both texts into news to reduce the computational complexity. Similarly, to make a fair comparison, we also normalize the inconsistency.

 The objective function of inferring clustering confidence matrices that can minimize the normalized inconsistency can be formulated as follows:

\begin{center}
\begin{equation}
\min_{H^{(1)},H^{(2)}}\frac{\|\bar{H}^{(1)}(\bar{H}^{(1)})^T-\bar{H}^{(2)}(\bar{H}^{(2)})^T\|_F^2}{|\mathcal{R}|(|\mathcal{R}|-1)}.
\label{eq:eq9}
\end{equation}
\end{center}

Figure \ref{fig:fig4} gives an example of the basic steps in the calculation of inconsistency and normalized inconsistency. The anchored news $t_1^{(2)},t_3^{(2)}$ are in the same cluster initially, but the corresponding tweets $t_1^{(1)},t_3^{(1)}$ and $t_2^{(1)},t_4^{(1)}$ are  in different clusters.  We calculate the inconsistency and normalize inconsistency of the figure by formula (\ref{eq:eq9}) with $|\mathcal{R}| = 4$. In this case, the inconsistency between the two clustering results is 16, while the normalized inconsistency is 1.33.

To get a consensus clustering, we need to minimize the inconsistency according to the above inconsistency definition. Meanwhile, confidence matrices $\mathbf{H}^{(1)}$  and $\mathbf{H}^{(2)}$  are of different dimensions, e.g., $(T^{(1,2)})^T\mathbf{H}^{(1)} \in \mathbb{R}^{3 \times 2}$ and $(T^{(2,1)})^T\mathbf{H}^{(2)} \in \mathbb{R}^{4 \times 2}$. To represent the inconsistency with the clustering confidence matrices, we need to further accommodate the dimensions of different pruned clustering confidence matrices. It can be achieved by multiplying one pruned clustering confidence matrix with the corresponding anchor transition matrix again, which only needs to adjust the matrix dimensions. Let $\bar{\mathbf{H}}^{(1)}=(T^{(1,2)})^T\mathbf{H}^{(1)}$ and $\bar{\mathbf{H}}^{(2)}=(T^{(1,2)})^T(T^{(2,1)})^T\mathbf{H}^{(2)}$. In the example, we can represent the clustering inconsistency to be $||\mathbf{H}^{(1)}(\mathbf{H}^{(1)})^T- (\mathbf{H}^{(2)})(\mathbf{H}^{(2)})^T||_F^2 =0$, where matrix $\mathbf{H}(\bar{\mathbf{H}})^T$ indicates whether pairs of anchor users are in the same cluster or not.

\subsection{Balanced Clustering on HINs}

We consider the unique characteristics as well as the connections of the comparative texts simultaneously for clustering and find out the corresponding information to mutually refine the clusterings. Based on the above introduction on our model, we define a complete form of the objective function as follows:

\begin{center}
\begin{equation}
\begin{split}
\min_{\mathbf{H}^{(1)},\mathbf{H}^{(2)}} \alpha & \cdot Tr((\mathbf{H}^{(1)})^T\mathbf{L}^{(1)}\mathbf{H}^{(1)})+ \beta \cdot Tr((\mathbf{H}^{(2)})^T\mathbf{L}^{(2)}(\mathbf{H}^{(2)}))\\
+ & \theta \cdot \frac{||\mathbf{H}^{(1)}(\mathbf{H}^{(1)})^T- (\mathbf{H}^{(2)})(\mathbf{H}^{(2)})^T||_F^2}{|\mathcal{R}|(|\mathcal{R}|-1)},\\
s.t. & (\mathbf{H}^{(1)})^T\mathbf{D}^{(1)}\mathbf{H}^{(1)} = \mathbf{I} ; \; (\mathbf{H}^{(2)})^T\mathbf{D}^{(2)}\mathbf{H}^{(2)} = \mathbf{I},
\end{split}
\label{eq:eq10}
\end{equation}
\end{center}

\noindent where $\alpha$, $\beta$, and $\theta$  are the parameters denoting the weights of news clustering, tweet clustering, and the inconsistency between them. Since the two types of texts are both textual documents, in this paper, we set both $\alpha$  and $\beta$ as 1; the constraints $\mathbf{L}^{(1)}$ and $\mathbf{L}^{(2)}$ are the Laplacian matrices as we defined in Formula (\ref{eq:eq3}) corresponding to tweets $T^{(1)}$ and news $T^{(2)}$, respectively.

Unfortunately, the objective function (\ref{eq:eq10}) is non-convex and the orthogonal constraints are expensive to preserve in calculation. However, since the constraints are orthogonal, by substituting $(\mathbf{D}^{(1)})^{\frac{1}{2}}\mathbf{H}^{(1)}$ and $(\mathbf{D}^{(2)})^{\frac{1}{2}}\mathbf{H}^{(2)}$ with $\mathbf{X}^{(1)}$ and $\mathbf{X}^{(2)}$, we can transform the objective function into a standard form of problem solvable with the method proposed in \cite{wen2013feasible}.

\begin{equation}
\begin{split}
\min_{\mathbf{X}^{(1)},\mathbf{X}^{(2)}}  & Tr((\mathbf{X}^{(1)})^T\tilde{L}^{(1)}\mathbf{X}^{(1)})+ Tr((\mathbf{X}^{(2)})^T\tilde{L}^{(2)}\mathbf{X}^{(2)})\\
+ & \theta \cdot \frac{\|\mathbf{\tilde{T}}^{(1)}\mathbf{X}^{(1)}(\mathbf{\tilde{T}}^{(1)}\mathbf{X}^{(1)})^T-\mathbf{\tilde{T}}^{(2)}\mathbf{X}^{(2)}(\mathbf{\tilde{T}}^{(2)}\mathbf{X}^{(2)})^T\|_F^2}{|\mathcal{R}|(|\mathcal{R}|-1)}\\
s.t. & (\mathbf{X}^{(1)})^T\mathbf{X}^{(1)} = \mathbf{I}; (\mathbf{X}^{(2)})^T\mathbf{X}^{(2)} = \mathbf{I}
\end{split}
\label{eq:eq11}
\end{equation}

Through a comparison between formulas  (\ref{eq:eq10}) and  (\ref{eq:eq11}), we can get the variables $\tilde{L}^{(1)}=((D^{(1)})^{-1/2})^TL^{(1)}((D^{(1)}) ^{-1/2})$, $\tilde{L}^{(2)}=((D^{(2)})^{-1/2})^TL^{(2)}((D^{(2)}) ^{-1/2})$, $\tilde{T}^{(1)}=(T^{(1,2)})^T(D^{(1)})^{-1/2}$, and $\tilde{T}^{(2)}=(T^{(2,1)})^T(D^{(2)})^{-1/2}$.

We utilize the feasible method proposed in \cite{wen2013feasible} for alternatively solving $\mathbf{X}^{(1)}$ and $\mathbf{X}^{(2)}$, and propose a double iteration process with the constraint-preserving update scheme. The scheme alternatively updates one variable (e.g. $\mathbf{X}^{(1)}$)  while fixing the other variable (e.g. $\mathbf{X}^{(2)}$),  by the curvilinear search with Barzilai-Borwein step until convergence. For example, when  $\mathbf{X}^{(1)}$ or $\mathbf{X}^{(2)}$ is fixed, we can simplify the objective function as:

\begin{center}
\begin{equation}
\begin{split}
\min_{\mathbf{X}}\mathcal{F}(\mathbf{X})=Tr((\mathbf{X})^T\tilde{L}\mathbf{X})+ & C + \theta \cdot \frac{\|\mathbf{\tilde{T}}\mathbf{X}(\mathbf{\tilde{T}}\mathbf{X})^T-B\|_F^2}{|\mathcal{R}|(|\mathcal{R}|-1)}\\
s.t. (\mathbf{X})^T\mathbf{X} & =\mathbf{I}
\end{split}
\label{eq:eq12}
\end{equation}
\end{center}

\noindent where $\mathbf{X} = \mathbf{X}^{(1)}$ or $\mathbf{X}^{(2)}$, and terms $C$ and $B$ can be translated from Formula (\ref{eq:eq10}). $\mathcal{F}(\mathbf{X})$  is the objective function, which can be solved with the curvilinear search using Barzilai-Borwein (BB) step method proposed in \cite{wen2013feasible} until convergence. Since BB step method can not guarantee convergence in every step, we apply the strategy proposed in \cite{zhang2004nonmonotone} to generate next points iteratively with a guarantee of convergence.

\renewcommand{\algorithmicrequire}{\textbf{Input:}}
\renewcommand{\algorithmicensure}{\textbf{Output:}}
\begin{algorithm}
  \caption{The core algorithm of HINT}
  \label{alg:alg1}
  \begin{algorithmic}[1]
  \Require tweet $T^{(1)}$, news $T^{(2)}$, anchor text $R$\
           number of clusters of $T^{(1)}$, and $T^{(2)}$ :  $k^{(1)}$ and $k^{(2)}$;\
           HINTS Sim matrices weight : $\omega$;\
           function $\mathcal{F}$ and consensus term weight $\theta$
\Ensure $\mathbf{H}^{(1)}$, $\mathbf{H}^{(2)}$
  \State Calculate HINTS matrices, $\mathbf{S}_i^{(1)}$ and $\mathbf{S}_i^{(2)}$ \label{alg:step1}
  \State $\mathbf{S}^{(1)}= \Sigma_i \omega_iS_i^{(1)}$,
  $\mathbf{S}^{(2)}= \Sigma_i \omega_iS_i^{(2)}$
  \State Initialize $\mathbf{X}^{(1)}$ and $\mathbf{X}^{(2)}$ with K-means clustering results on $\mathbf{S}^{(1)}$ and $\mathbf{S}^{(1)}$ \label{alg:step3}
  \State converge = false
  \While {converge==false} \label{alg:step5}
      \State  $\mathbf{X} = \mathbf{X}^{(1)}$
      \State  Search next $\mathbf{X}$ by step in \cite{wen2013feasible}. \label{alg:step7}
      \State  $\mathbf{X} = \mathbf{X}^{(2)}$
      \State  Repeat Step 7.
        \If{$\mathcal{F}(\mathbf{X}^{(1)})$ and $\mathcal{F}(\mathbf{X}^{(2)})$ both converge}
        \State converge = true
    \EndIf
  \EndWhile \label{alg:step13}
  \State \textbf{return} $\mathbf{H}^{(1)}=((\mathbf{D}^{(1)}))^T\mathbf{X}^{(1)}$,$\mathbf{H}^{(2)}=((\mathbf{D}^{(2)}))^T\mathbf{X}^{(2)}$
\end{algorithmic}
\end{algorithm}

\subsection{Complexity Analysis}

The core algorithm of HINT is presented in Algorithm \ref{alg:alg1}. Steps \ref{alg:step1}-\ref{alg:step3} of the algorithm make a preparation on the balanced clustering, and Steps \ref{alg:step5}-\ref{alg:step13} conduct the balanced clustering to obtain $\mathbf{H^{(1)}}$ and $\mathbf{H^{(2)}}$.

Since our algorithm contains two parts of the searching for the optimized solution, the complexity of our algorithm comes from the two parts.  For each optimization step, we use the BB step to search for the right $\tau$. Thus the complexity of this part is $O(\tau_1+\tau_2)$. The searching process depends on the maximum iteration step of the BB method. According to Wen et al, the method is guaranteed to converge, which depends on parameter $h$ which is the smallest integer satisfying $\mathcal{F}(Y_k(\tau_k)) \le C_k + \rho_1\tau_k\mathcal{F}'(Y_k(0))$. Therefore, the total time complexity of our algorithm is $O(\tau_1+\tau_2)(h_1+h_2)$. In fact, if we set the maximum iteration times of the BB method as $k$, The complexity of our algorithm will be $O(k(\tau))$. Note that the algorithm usually converges very quickly due to the application of Wen et al's new method \cite{wen2013feasible}.

\section{Experimental Results}\label{sec:sec5}

\subsection{Dataset and Baselines}

We evaluate the proposed model on the following three datasets.

\textbf{Dataset 1.} This dataset is collected by Guo et. al \cite{guo2013linking}, and contains tweets spanning over 18 days. Each tweet contains a URL linking to a news article of CNN or NYTIMES published in the same time period.

\textbf{Dataset 2.}  The original tweet corpora was crawled from the over 4 million followers of Hillary Clinton. We keep the tweets containing news URLs, and randomly select other tweets posted from June 1st, 2015 to June 7th, 2015 to make up the tweet part. The news part was crawled based on the URLs in the tweets and covered over 20 major news sites.

\textbf{Dataset 3.} Similar to Dataset 2, the tweet part consists of the tweets with URLs and we randomly select part of the data posted from June 25th, 2015 to July 2nd, 2015. The news part was crawled based on the URLs in the tweets that covered over 20 news sites.

We first preprocess the data by removing stop words, extracting hashtags, mentions, keywords, and named entities on tweets, and also named entities and keywords on news. We solve the short URLs in tweet to expanded identical URLs, construct the substitution of out-of-vocabulary words, and move out long-tailed users, tweets. Following \cite{wang2015incorporating}, we use three types of entities, the person (P), organization (O) and location (L) in tweets and news. After data preprocessing, the three corpora associated with properties are obtained and the detailed dataset description is summarized in Table \ref{tab:tab1}.

We have three annotators to annotate the data as the ground truth, and the consensus of their annotation is 98.5\%. Note that the annotators take consideration of both types of texts when they cluster each type of text, and thus the final annotation result also incorporates the semantic correlations between tweets and news.

\textbf{Baseline Methods.} We examine the effectiveness of the proposed HINT by comparing it with the following text clustering baselines.

$\bullet$ \textit{K-Means.} We use the classical method K-Means to separately cluster the two types of texts through their similarity matrices as the first baseline.

$\bullet$ \textit{K-Medoids.} The K-medoids method is an extension of the K-means algorithm by using the \textit{kernel trick} \cite{Aggarwal2015Data}. K-medoids is also a widely used clustering method and usually achieves more promising clustering results compared with K-Means. 

$\bullet$ \textit{RankClus.} RankClus is a clustering method utilizes the ranking result as the feature to improve clustering results \cite{Sun2009RankClus}. Since it can be only applied on bi-typed information networks, we preserve the tweet-news part in our constructed heterogeneous network and ignore the other parts. The calculation of the correlation scores is similar to our proposed model.

$\bullet$ \textit{TF-IDF.} As a classical text representation method, TF-IDF takes into consideration of the term frequencies (TF) and inverse document frequency (IDF). Since TF-IDF does not fit for short text clustering \cite{jin2011transferring}, in this paper we employ TF-IDF (+ K-Means) to cluster tweets.

$\bullet$ \textit{SpecClus (Spectral Clustering).} Spectral clustering makes use of the spectrum (eigenvalues) of the similarity matrix of the data to perform dimensionality reduction before clustering \cite{Luxburg2007A}. Considering that our edges of HINs are weighted, we use the widely adopted normalized cuts method to perform spectral clustering for comparison \cite{Shi2000Normalized}. To be fair, we set the input of spectral clustering as the similarity matrices of both tweet and news texts and the same number of clusters $k$ used by HINT.

$\bullet$ \textit{WKF.} The Web-based Kernel Function (WKF) for measuring the similarity is an effective similarity measure proposed in \cite{sahami2006web}. WKF uses the semantic vectors associated with text snippets to represent them, thus it can  enrich the semantical space, and assist the calculation of short text similarities. In this paper, to be fair, we use the same query vectors generated from the news corpora and  use different settings of parameter $m$ to compare it with HINT in tweets.

$\bullet$ \textit{WTMF.} Weighted Textual Matrix Factorization  (WTMF) is a latent variable model for the task of measuring text similarity. According to \cite{guo2013linking}, WTMF is a state-of-the-art unsupervised model which outperforms Latent Semantic Analysis (LSA) and Latent Dirichlet Allocation (LDA) by a large margin on short text similarity

$\bullet$ \textit{WTMF-G.} WTMF on Graphs (WTMF-G) is an extended version of WTMF, which tightens the relationships of tweets and news through semantic latent vectors. Note that we employ the WTMF and WTMF-G with K-Means to cluster the tweet and news as baselines.

$\bullet$ \textit{Girvan-Newman (GN).} Girvan-Newman algorithm is one typical algorithm that uses betweenness to generate clusters \cite{Girvan2002Community}. Since this algorithm uses edge length $c_{ij}$ rather than the edge weight $w_{ij}$, we transform the edge weight to length through $c_{ij} = 1/w_{ij}$.

\subsection{Experiment Settings}

In this part, we first conduct parameter sensitivity analysis of HINT, and then introduce the baselines used in the following experiments. There are three groups of parameters in the HINT model: (1) the number of clusters for each data collection (we use different settings in the following experiments), (2) the weight of different features $\omega$, and (3) the consensus term weights $\theta$. For simplicity, we set all the  meta-path weight $\omega$ as a diagonal matrix with the diagonal vector of tweet meta-paths as $[\frac{1}{6}, \cdots,\frac{1}{6}]$, and that of news meta-paths as $[\frac{1}{4}, \cdots, \frac{1}{4}]$. Here, we set the hyper-parameter of the consensus term weight $\theta$ through the final clustering results.

\begin{table}
\centering
\small
\caption{Statistics of the Three Datasets}
\begin{tabular}{c|c c c c}
\hline
                                  & Objects             &Dataset 1    &Dataset 2   &Dataset 3\\
                                  \hline 
   \multirow{6}{*}{Tweets}        & \#Tweets            &  34,888     &  5,628     & 3,847     \\
                                  & \#Hashtag           &  15,471     &  1,602     & 1,526     \\
                                  & \#Mention           &  12,911     &  4,120     & 2,950     \\
                                  & \#Keyword           &  13,322     &  7,325     & 4,527     \\
                                  & \#HyperLink         &  2,710      &  4,345     & 3,784     \\
                                  & \#Entity            &  5,468      &  1,237     & 666      \\
                                  \hline 
   \multirow{2}{*}{Anchor text}   & \#Tweets            &  34,888     &  2,914     & 2,137      \\
                                  & \#News              &  12,704     &  3,653     & 2,825      \\
                                  \hline 
   \multirow{3}{*}{News}          & \#News              &  12,704     &  3,653     & 2,825     \\
                                  & \#Entity            &  8,325      &  8,696     & 18,696     \\
                                  & \#Keyword           &  16,210     &  17,475    & 21,475     \\
                                  \hline 
\end{tabular}
\label{tab:tab1}
\end{table}

\textbf{Setting Hyper-Parameter $\theta$.} In the mutual clustering of tweets and news, $\theta$ controls the refinement process of the mutual clustering. We use the anchor texts to find the best setting of the hyper-parameter $\theta$. We fix the values of the other parameters and set $\theta$ to different values for investigating the clustering results. A larger anchor texts rate in the final mutual clusterings means the better performance the algorithm achieving under the parameter setting of $\theta$. We randomly choose 80\% of the anchor texts as the training dataset and the remaining 20\% as testing dataset. Then we tune the parameter $\theta$ based on the anchor rate of the final mutual clustering results. If the anchored texts are in the correlated clusters, we denote it as correct; otherwise, we denote it as incorrect. We initialize the hyper-parameter $\theta$ as 0.8 and find that the anchor text rate of the testing dataset keeps stable when $\theta$ falls in the range of $[0.5,1.5]$.

\begin{figure}
\begin{minipage}[t]{0.5\linewidth}
\centering
\includegraphics[width=2.4in]{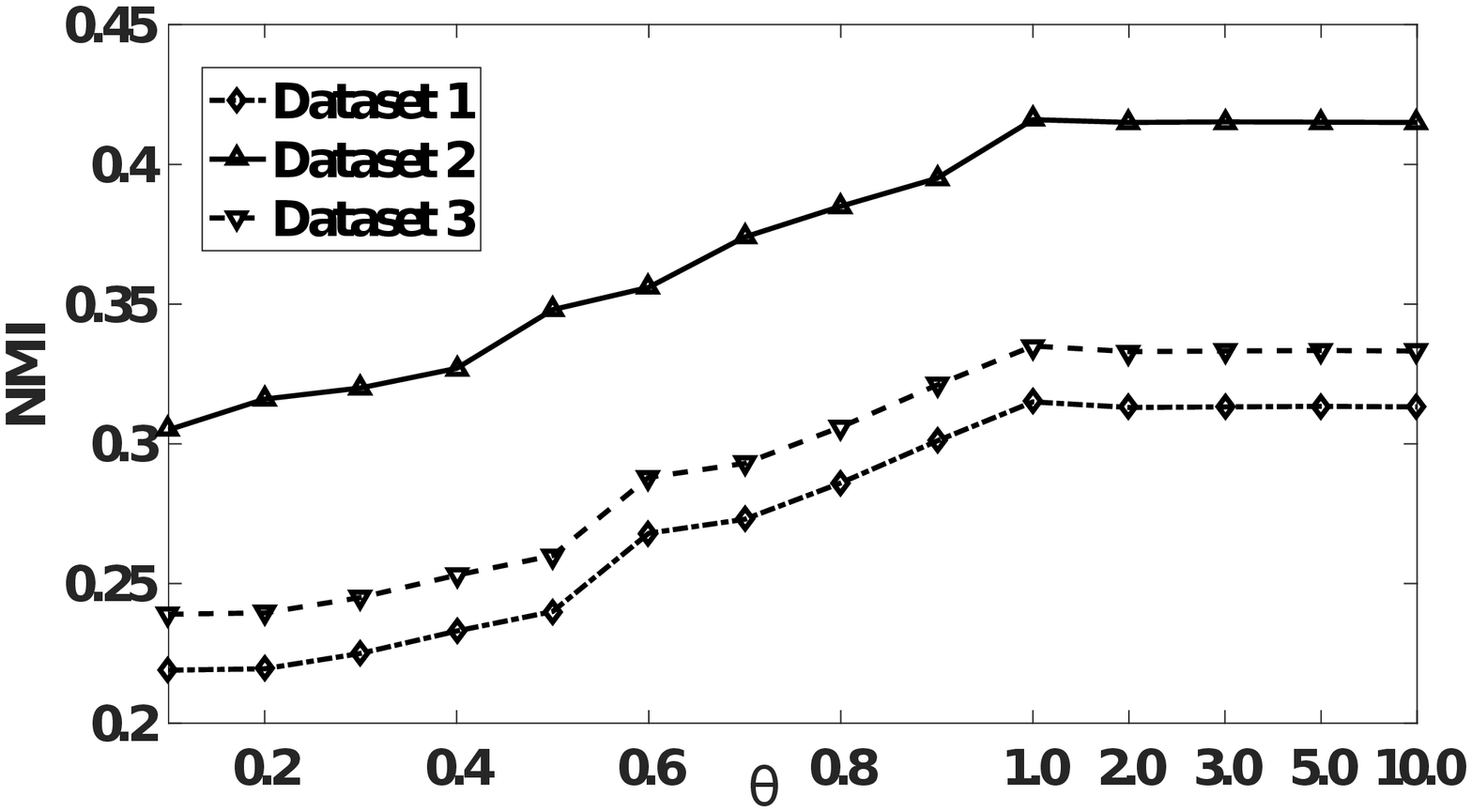}
\caption{Sensitivity analysis of parameter $\theta$.}
\label{fig:side:a}
\end{minipage}%
\begin{minipage}[t]{0.5\linewidth}
\centering
\includegraphics[width=2.7in]{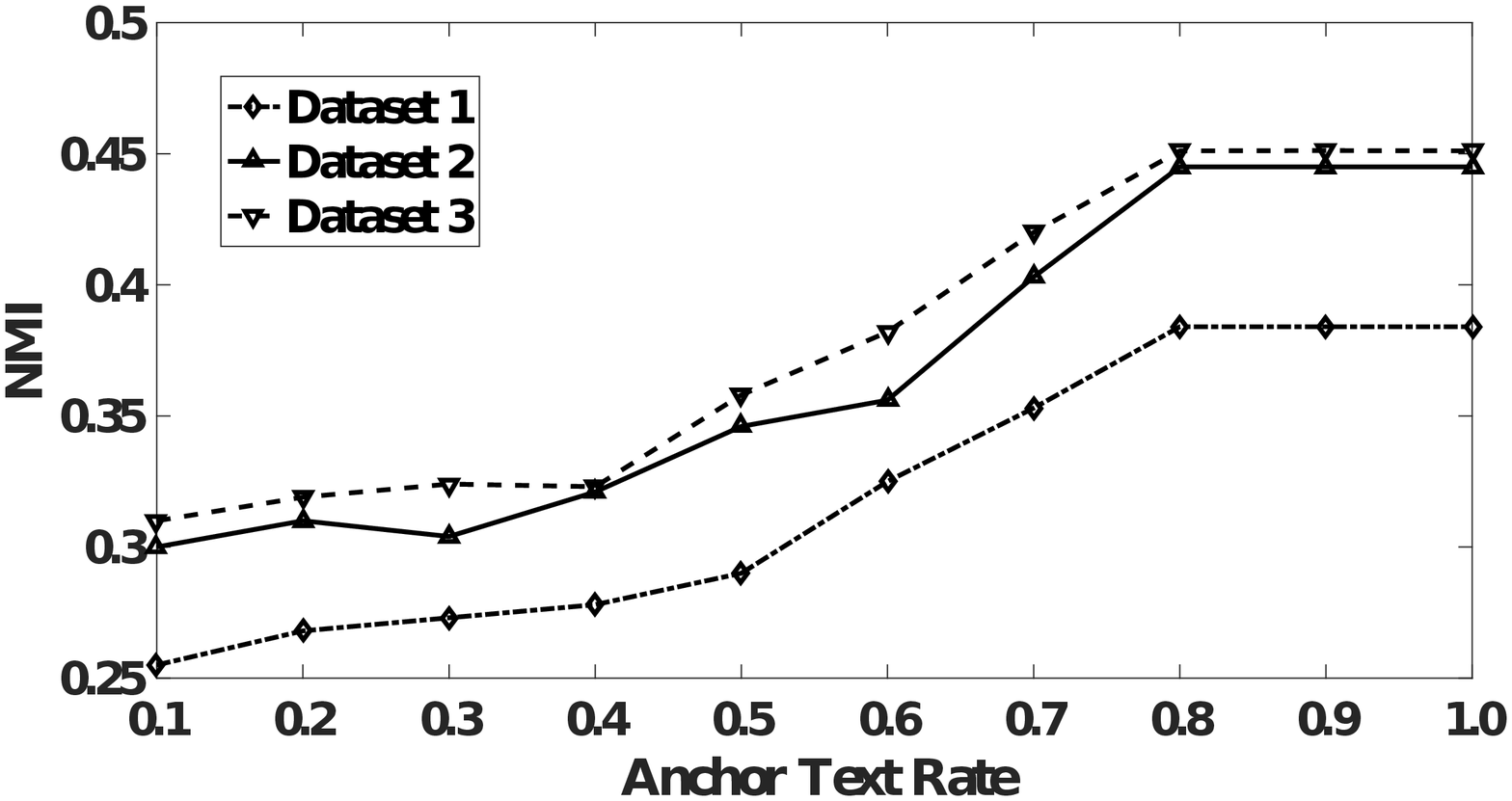}
\caption{NMI \textit{vs} different anchored text rates.}
\label{fig:side:b}
\end{minipage}
\end{figure}

\textbf{Analysis on Hyper-Parameter $\theta$.} Considering that the parameter $\theta$ has great effect on the final clustering results, we further determine the best choice of  $\theta$ by investigating its effect on the final clustering performance. As shown in Figure \ref{fig:side:a}, one can see that $\theta$ has a significant effect to the final clustering results on both collections. When $\theta$ is relatively small, the penalty of inconsistency is insignificant, and the performance improvement on mutual clustering is not remarkable. When $\theta$ becomes larger, the clustering result of tweets becomes better with smaller variation, and the mean results of news clustering are stable. The results indicate that a larger $\theta$ makes the clustering more robust. The clustering results remain stable when $\theta$ is larger than 1.0. In this paper, we set $\theta$ as 1.0 in the following experiments.

\textbf{Evaluation metrics and other settings.} We apply two widely used evaluation metrics, NMI and F-score to evaluate the performance of clustering results. Here we use F1-score to evaluate the clustering results of news and tweets respectively. All the results of each method under different settings are averaged over 100 runs for every dataset.

\subsection{Experimental Result Analysis}

\subsubsection{Analysis on the Clustering Results over the Comparative Texts of News and Tweets}

\newcommand{\tabincell}[2]{\begin{tabular}{@{}#1@{}}#2\end{tabular}}

\begin{table}
\centering
\small
\caption{NMI and F-score of various methods on the three datasets}
\begin{tabular}{c|c|c|c|c|c|c|c}
\cline{1-8}
\multicolumn{2}{c}{}              &\multicolumn{2}{|c|}{$k^{(1)}=k^{(2)}=10$}   &\multicolumn{2}{|c|}{$k^{(1)}=k^{(2)}=20$}   &\multicolumn{2}{|c|}{$k^{(1)}=k^{(2)}=30$} \\
\cline{1-8}
\multicolumn{2}{c|}{}           &NMI    &F-1    &NMI    &F-1     &NMI   &F-1   \\
\cline{1-8}
\multicolumn{8}{|c|}{Dataset 1}        \\ \cline{1-8}
   \multirow{10}{*}{T}  & K-Means        &0.274    &0.361   &0.116   &0.492    &0.117   &0.480 \\
                       & K-Medoids        &0.281    &0.365   &0.117   &0.504    &0.121   &0.493  \\
                       & RankClus       &0.142    &0.204   &0.059   &0.203    &0.056   &0.213 \\
                       & SpecClus       &0.328    &0.579   &0.294   &0.647    &0.216   &0.704\\
                       & TKF            &0.268    &0.349   &0.109   &0.468    &0.106  &0.468 \\
                       & WTMF           &0.301    &0.601   &0.302   &0.584    &0.205  &0.709 \\
                       & WTMFG          &0.356    &0.634   &0.341   &0.596    &0.218  &0.723 \\
                       & GN        &0.318    &0.509   &0.308   &0.492    &0.213   &0.719  \\
                       & HINT   &\textbf{0.420}  &\textbf{0.661}   &\textbf{0.394}  & \textbf{0.687} &\textbf{0.348}   &\textbf{0.826} \\
                       \cline{1-8}
   \multirow{10}{*}{N}  & K-Means    &0.423   &0.705   &0.412   &0.696    &0.276  &0.589\\
                       & K-Medoids        &0.426    &0.706   &0.431   &0.643    &0.281   &0.607  \\
                       & RankClus   &0.205   &0.341   &0.204   &0.377    &0.142  &0.275\\
                       & SpecClus   &0.463   &0.817   &0.434   &0.799    &0.327  &0.759\\
                       & TF-IDF     &0.392   &0.683   &0.409   &0.684    &0.274  &0.563 \\
                       & WTMF       &0.401   &0.631   &0.421   &0.705    &0.206  &0.659 \\
                       & WTMFG      &0.426   &0.620   &0.406   &0.776    &0.204  &0.658 \\
                       & GN        &0.403    &0.619   &0.401   &0.735    &0.298   &0.674  \\
                       & HINT   &\textbf{0.494}  &\textbf{0.821}   &\textbf{0.458}  &\textbf{0.792}  &\textbf{0.351}   &\textbf{0.777} \\
                       \cline{1-8}
                       \multicolumn{8}{|c|}{Dataset 2}        \\ \cline{1-8}
   \multirow{10}{*}{T}  & K-Means    &0.117  &0.718    &0.123    &0.492    &0.121   &0.275 \\
                       & K-Medoids        &0.193    &0.729   &0.163   &0.511    &0.199   &0.309  \\
                       & RankClus   &0.049  &0.328    &0.058    &0.235    &0.062   &0.141 \\
                       & SpecClus   &0.317  &0.721    &0.280    &0.782    &0.248   &0.801\\
                       & TKF        &0.098   &0.714   &0.103   &0.478    &0.103  &0.259 \\
                       & WTMF       &0.328   &0.726   &0.206   &0.589    &0.203  &0.694 \\
                       & WTMFG     &0.326   &  0.712 &0.271   &0.604    &0.206  &0.706 \\
                       & GN        &0.103    &0.411   &0.275   &0.597    &0.223   &0.472  \\
                       & HINT   &\textbf{0.355}  &\textbf{0.866}   &\textbf{0.329}  & \textbf{0.862} &\textbf{0.392}   &\textbf{0.849} \\
                       \cline{1-8}
   \multirow{10}{*}{N}  & K-Means    &0.296   &0.614   &0.305   &0.629    &0.276  &0.583 \\
                       & K-Medoids        &0.298    &0.612   &0.307   &0.626    &0.281   &0.594  \\
                       & RankClus   &0.145   &0.301   &0.161   &0.321    &0.145  &0.258 \\
                       & SpecClus   &0.307   &0.641   &0.313   &0.618    &0.306  &0.621\\
                       & TF-IDF &0.285   &0.596   &0.293   &0.609    &0.261  &0.548 \\
                       & WTMF   &0.296   &0.583   &0.269   &0.602    &0.265  &0.613 \\
                       & WTMFG  &0.305   &0.608   &0.309   &0.608    &0.286  &0.619 \\
                       & GN        &0.300    &0.507   &0.298   &0.603    &0.278   &0.622  \\
                       & HINT   &\textbf{0.310}  &\textbf{0.642}   &\textbf{0.316}  &\textbf{0.633}  &\textbf{0.310}   &\textbf{0.635} \\
                       \cline{1-8}
                      \multicolumn{8}{|c|}{Dataset 3}        \\ \cline{1-8}
   \multirow{10}{*}{T}  & K-Means    &0.103  &0.594   &0.141   &0.521    &0.105  &0.321  \\
                       & K-Medoids        &0.109    &0.599   &0.143   &0.509    &0.113   &0.394  \\
                       & RankClus   &0.056  &0.315   &0.074   &0.241    &0.073  &0.138 \\
                       & SpecClus   &0.274  &0.794   &0.271   &0.603    &0.273  &0.684\\
                       & TKF    &0.089   &0.581   &0.127   &0.504    &0.013  &0.312 \\
                       & WTMF   &0.253   &0.653   &0.256   &0.542    &0.216  &0.598 \\
                      & WTMFG  &0.261   &0.792   &0.268   &0.549    &0.218  &0.629 \\
                       & GN        &0.249    &0.707   &0.215   &0.552    &0.253   &0.639  \\
                       & HINT   &\textbf{0.293}  &\textbf{0.806}   &\textbf{0.289}  & \textbf{0.628} &\textbf{0.316}   &\textbf{0.762} \\
                       \cline{1-8}
   \multirow{10}{*}{N}  & K-Means    &0.248   &0.542   &0.282   &0.686    &0.262  &0.599 \\
                       & K-Medoids        &0.257    &0.558   &0.290   &0.701    &0.263   &0.615  \\
                       & RankClus   &0.116   &0.256   &0.237   &0.330    &0.135  &0.305 \\
                       & SpecClus   &0.314   &0.649   &0.372   &0.714    &0.314  &0.671\\
                       & TF-IDF     &0.236   &0.503   &0.268   &0.642    &0.240  &0.549 \\
                       & WTMF       &0.284   &0.642   &0.361   &0.708    &0.300  &0.609 \\
                       & WTMFG      &0.301   &0.650   &0.374   &0.706    &0.308  &0.618 \\
                       & GN        &0.195    &0.470   &0.237   &0.601    &0.218   &0.446  \\
                       & HINT       &\textbf{0.329}  &\textbf{0.653}   &\textbf{0.389}  &\textbf{0.728}  &\textbf{0.319}   &\textbf{0.684} \\
                       \cline{1-8}
\end{tabular}
\label{tab:tab2}
\end{table}

Table \ref{tab:tab2} shows the NMI and F-score results of each method on the three datasets. Here, we set the cluster numbers of news and tweets in each dataset as equal as 10, 20 and 30 respectively. One can see that HINT outperforms all the baselines on both the tweet clustering (as shown in the row labeled ``T'') and the news clustering (as shown in the row labeled ``N''). Specifically, we have the investigations of the experimental results as follows:

$\bullet$  In all cases of the three dataset, HINT outperforms the K-means and K-Medoids consistently. As all of the three approaches use the same similarity matrices of tweet and news, we can conclude that the constraints from the counterpart collections will significantly improve the clustering performance. Moreover, we also compare our algorithm with RankClus which mainly focuses on exploring the correlation between tweets and news while ignores the information within tweet and new collections. The performance of HINT outperforms that of RankClus significantly, which implies that relying on the correlation information only can not achieve a promising performance. The above two experiments prove that HINT is more effective to explore information both within and cross the two collections.

$\bullet$  Compared to the methods which can be only utilized to cluster one type of data collection (TKF and TF-IDF), the mutual clustering methods HINT and WTMF are generally more effective, which indicates that mutual clustering does improve the clustering performances. Compared to the classical method TF-IDF, HINT improves the NMI by 8\%\text{-}45\% and F-score by 4\%\text{-}38\%. Such a result indicates that using HINs to represent the heterogenous texts can significantly improve the clustering performance.

$\bullet$  As for the short text clustering, HINT achieves a much better performance compared to the baselines. On average, HINT outperforms WKF over tweet collections by about 20\% in NMI and about 20\%\text{-}80\% in F-score respectively. The results indicate that transferring information between the two comparative texts improves the clustering results on both collections. Compared to WTMF, HINT outperforms them by about 30\% on NMI and about 15\% on F-score. Generally, HINT can more effectively use the information from both the tweet and news, and the promising performance indicates that HINT can better organize the correlated comparative texts.

$\bullet$ As for the graph based clustering methods, HINT achieves a better performance compared with the baselines. For spectral clustering, we use the same similarity matrices of tweets and news and the same number of clusters for both HINT and SpecClus. From Table \ref{tab:tab2} one can see that the spectral clustering method performs better than other methods in news clustering rather than tweet clustering. However, our proposed method HINT  performs better both in news and tweet clusterings. Comparing with spectral clustering which runs the process separately, we summarize that HINT achieves better performance due to the advantage of transition matrices to transmit information between the confidence matrices. As for the classical approache GN, although it achieves better performance than other distance based approaches, especially on the tweet clustering, it is still inferior to our proposed HINT method on all of the three datasets.

\subsubsection{Analysis on the Correlations of the Clusters of Comparative Texts}

In this part, we evaluate the correlations of the tweet and news clusters discovered by various methods. As one of the motivations of this study, we hope that the cluster correlations of the comparative texts can be high. Therefore, we employ conditional entropy to measure the correlation of comparative texts since it is widely used as correlation index between two sets. For simplicity, we selects the tweets and news that can form anchor texts to investigate the correlations between different clustering results.

We use conditional entropy as a metric to measure the correlations of clusters. Conditional  entropy quantifies the amount of information needed to describe the outcome of random variable $Y$ given that the value of another random variable $X$ is known. The entropy of $Y$ conditioned on $X$ is defined as:

\begin{center}
\begin{equation}
H(Y|X)= \sum_{x \in X, y \in Y} p(x,y) log \frac{p(x)}{p(x,y)},
\label{eq:eq13}
\end{equation}
\end{center}

\noindent where $p(x,y)$ is the probability of common elements between $X$ and $Y$, and $p(x)$ denotes the probability of $P(X=x)$.

In this paper, we use matrices to denote the clustering results as the same form as  the confidence matrices $H^{(1)},H^{(2)}$ as illustrated in the bottom left part of Figure \ref{fig:fig4}. We transfer the news clustering information matrix $H^{(2)}$ to tweet collection by $T^{(2,1)}H^{(2)}$, and compare $T^{(2,1)}H^{(2)}$ with $H^{(1)}$ to investigate the correlations of the tweet and news clusters.

\begin{figure}
\centering
\epsfig{file=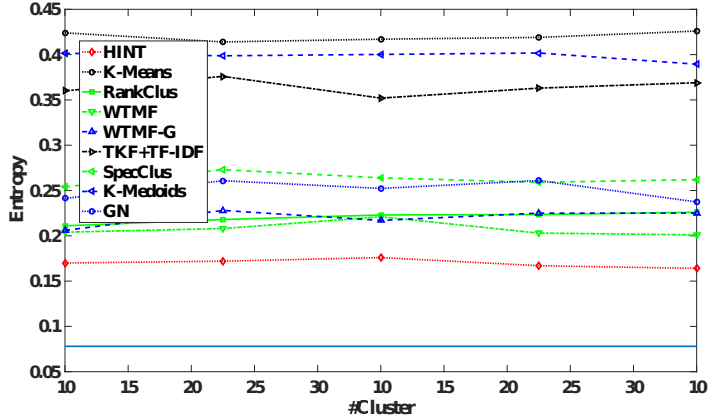, height=2.5in}
\caption{The conditional entropy of the tweet and news clusters on dataset 2.}
\label{fig:fig6}
\end{figure}

In our case, $H^{(1)}$ can be considered as $X$ of formula (\ref{eq:eq13}), and $T^{(2,1)}H^{(2)}$ can be considered as $Y$. The common elements probabilities $p(x,y)$ are obtained by $max P(X,Y=y)$. The smaller the conditional entropy is, the higher the correspondence is, and the higher correspondence indicates a larger correlation of the two clustering results. Figure \ref{fig:fig6} shows the correlations between the proposed methods and the ground truth on dataset 2. One can see that HINT shows a mush smaller conditional entropy and thus it achieves a much large correlation accordance compared with other methods, which means that it significantly outperforms other methods.

\begin{figure}
\centering
\epsfig{file=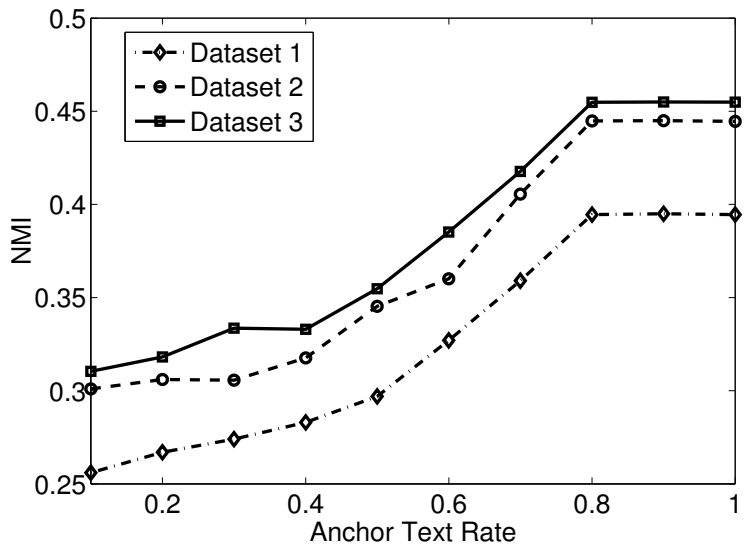, height=3in}
\caption{NMI \textit{vs} different anchored text rates.}
\label{fig:fig7}
\end{figure}

\subsection{Robustness Analysis}

As the performance of the proposed approach may be largely affected by the ``anchor texts,'' in this part, we examine the robustness of HINT by testing it with different rates of anchored texts. We use the anchored rate of tweet to indicate the anchor text rate, and study the robustness of HINT with various anchor text rates by investigating its performances with different rates of anchored tweet on the three datasets.

Figure \ref{fig:fig7} shows the effect of anchored tweet rate to the final tweet clustering results. One can see that the anchored tweet rate (the rate of the number anchored tweets in all used tweets) has a remarkable effect on the final clustering results. Initially, when the rate of anchored tweet is relatively small, the value of NMI is low. As the rate grows larger, the performance rises quickly. However, when it reaches 0.8, the NMI of the tweet clustering result reaches the peak value 0.6, and then the results become stable.

\subsection{Case study}

We give a case study on dataset 3 to further illustrate the effectiveness of HINT. We set parameter $\theta$ as 1, and set the cluster numbers of both tweets and news as 20. We use a simple rule (the anchored tweets $> 80$\%) to correlate the clusters of the two comparative texts obtained by HINT. Finally eight highly correlated clusters are discovered, and we show two of them in Table \ref{tab:tab3}. The two clusters are related to the topic of LGBT movement and climate change during this period. On June 27th, 2015 the verdict of the Supreme Court of U.S. granted the equal marriage rights. There were lots of people discussing about such an issue online and the news agencies also published a large amount of reports on it. During the same period, the topic about the Climate Change Conference was being heatedly discussed online, and reported by the journalists.

Table \ref{tab:tab3} illustrates the representative tweets and news titles in that period. From the table, one can learn that the first cluster denoted by $C1$ is about the LGBT movement, while the second cluster denoted by $C2$ is about the topic of climate change. The tweet clusters are semantically related to the news clusters respectively.  Moreover, although the two parts are about the same issue respectively, they provide us with information from different aspects. Taking the topic of LGBT as an example, the news articles are talking about the reactions from various organizations like the Republican Party and LGBT community, while the tweets contain more personal views from users, and the views are usually direct and emotional. Summarily, by combining the clustering results of the two text corpus, we can obtain more comprehensive information from different perspectives.

\begin{table}
\centering
\small
\caption{A case study of HINT}
\begin{tabular}{|l|l|l|} \cline{1-3}
   \multicolumn{1}{|c|}{}         &  \multicolumn{1}{|c|}{Tweets}   & \multicolumn{1}{|c|}{Titles of news articles}
     \\  \cline{1-3}
\multirow{10}{*}{C1}   & \tabincell{l}{Congratulations to on coming out, \\ on marriage plans. Who knew romance \\ could blossom at events? }    & \tabincell{l}{Conservative Republicans question \\ what's next after gay marriage ruling}    \\ \cline{2-3}
      & \tabincell{l}{The Greens will support every vote, \\ every time. But a bill the whole parliament \\ can own has the best chance of success. }    & \tabincell{l}{Anyone in any loving relationship \\ should get the legal benefits of marriage}  \\ \cline{2-3}
      & \tabincell{l}{RT Every day we see so much trouble \\ in the world. With the decision, \\ I am proud to celebrate that in America. }       &  \tabincell{l}{LGBT Latinos Despite Challenges \\ Greater Empowerment Progress}\\\cline{2-3}
      & \tabincell{l}{RT Today, love is set free. People of \\ absolute courage risked their lives, jobs.\\ reputations to help make this happen. }     &  \tabincell{l}{Unscripted Television Led the Charge\\ in Embracing LGBT Community  } \\\cline{2-3}
      & \tabincell{l}{One big SCOTUS decision went for will\\ the other shoe drop with legal gay marriage\\ across the country? Here's hoping.}     &    \tabincell{l}{U of S houses one of the largest \\ LGBT archives in the country } \\
      \cline{1-3}\cline{1-3}
\multirow{10}{*}{C2}   &  \tabincell{l}{ Hague judgement could inspire \\ a global civil movement via\\ http://t.co/wYt45kTxTB         }  &   \tabincell{l}{ UN climate talks moving at snails \\ pace says Ban Kimoon }  \\\cline{2-3}
            & \tabincell{l}{RT What does the Hague judgement mean? \\ Here's what thinks: http://t.co/pHlwcJqckw    }       &     \tabincell{l}{ Redford Time to step up game \\ on climate change } \\\cline{2-3}
            & \tabincell{l}{For gas to be part of the solution to \\and emissions we must manage emissions\\ - can/will we? }    &    \tabincell{l}{ On climate change Hispanic Catholics \\ hear popes message  and its personal }    \\\cline{2-3}
            & \tabincell{l}{Court ordered government to \\cut to protect citizens from }     &        \tabincell{l}{ Pope Francis recruits Naomi Klein in \\  climate change battle }\\\cline{2-3}
            & \tabincell{l}{Catholic Latinos care more about climate\\ change than their white counterparts. by\\ http://t.co/iA1VPoAxyL }    &    \tabincell{l}{Barack Obama sets sizzling climate \\action pace in push to leave legacy } \\
           \cline{1-3}
\end{tabular}
\label{tab:tab3}
\end{table}

\subsection{Experiment Results Discussion}\label{sec:sec6-2}

In this subsection we would like to discuss our approach and the experimental results. Following the traditional experimental settings in this area, the numbers of clusters $k_1$ and $k_2$ are usually chosen by users \cite{Han2011Data,Luxburg2007A}. One may wonder why we let $k_1$ be equal to $k_2$. In fact, if we keep $k_1$ remain the same with different settings of $k_2$  or vice versa, the mutual clustering results will not change too much. Therefore, in this paper, we only report the results when $k_1$ and $k_2$ are the same for simplicity. In addition, the efficiency of our algorithm is comparable with the traditional methods like k-means and spectral clustering. Since our research focus is on improving the accuracy and F1-score of the mutual clustering on comparative texts, we do not conduct the experiments to show the efficiency of our method.

\section{Conclusions}\label{sec:sec7}

In this paper, we studied the novel problem of mutual clustering on comparative texts by taking tweets and news articles as an example. We proposed a framework HINT to address the mutual clustering problem. The proposed framework consists of three steps: (1) transforming the comparative texts into heterogeneous information networks, and connecting them through the anchor texts; (2) constructing the similarity matrices and transition matrix based on the constructed connected networks; and (3) proposing a balanced mutual clustering approach on the constructed matrices using a curvilinear search algorithm. We extensively evaluated the proposed model on three tweets-news comparative text datasets, and the experimental results have shown the effectiveness and robustness of the proposed framework.

\section{Acknowledgments}
This work is supported in part by the National Key R\&D Program of China (No: 2018YFB1003900), National Natural Science Foundation of China (Nos: 61602237, 61533019, 61672313), Natural Science Foundation of Jiangsu Province of China (No: BK20171420), Beijing Municipal Science \& Technology Commission (No: Z181100008918007), NSF through grants IIS-1526499, IIS-1763325, and CNS-1626432.

\label{lastpage}

\end{document}